# Content-based image retrieval system with most relevant features among wavelet and color features

Abdolreza Rashno
Department of Computer Engineering,
Lorestan University
Khorramabad, Iran
ar.rashno@gmail.com

Elyas Rashno
Department of computer engineering
Iran University of science and technology
Tehran, Iran
elyas.rashno@gmail.com

## Abstract

Content-based image retrieval (CBIR) has become one of the most important research directions in the domain of digital data management. In this paper, a new feature extraction schema including the norm of low frequency components in wavelet transformation and color features in RGB and HSV domains are proposed as representative feature vector for images in database followed by appropriate similarity measure for each feature type. In CBIR systems, retrieving results are so sensitive to image features. We address this problem with selection of most relevant features among complete feature set by ant colony optimization (ACO)-based feature selection which minimize the number of features as well as maximize F-measure in CBIR system. To evaluate the performance of our proposed CBIR system, it has been compared with three older proposed systems. Results show that the precision and recall of our proposed system are higher than older ones for the majority of image categories in Corel database.

*Key words*: Content-based image retrieval, wavelet transform, color, ant colony optimization, feature selection.

## 1. Introduction

Image retrieval is a task of retrieving relevant images in image databases which has an important role in different medical, economic and other fields recently [7-12, 21]. The first ideas were the retrieving of images based on text describers which were created by human, without considering their visual characteristics [10]. CBIR is a search method using the structural and conceptual characteristics of the image. Due to the increasing demand for optimum retrieving of images in large image data bases, research in the field of image retrieval has gained high attention. In general, image retrieval system is divided into three parts of user interface, data base and the processing section. One of the most important achievements of image retrieval is CBIR system. Many methods have been proposed for CBIR [7-10-26-51-56, 57]. Prior to this work, we have proposed CBIR systems based on color and texture features with feature selection methods and neutrosophic (NS) theory [43-45]. After that, we also applied NS theory in other applications [31-38]. NS theory also applied on speech processing and clustering applications [58-59].

One of the most significant parts in CBIR system is feature extraction [22, 57]. Color, texture and shape are three types of low-level features which are proposed for CBIR systems. The color is widely used in CBIR systems since its extraction is usually easy as well as its performance is relatively high for retrieving task [17-22-26, 34]. For shape features, a comprehensive and exact definition has not been presented yet. Generally, methods like aspect ratio, circularity, and Fourier descriptor are used for extracting the geometric shape features [8, 52]. Also, three methods include statistical, structural, and spectrum methods are used for describing the texture features [1, 29]. Furthermore, wavelet features are a type of texture features which have been used in many image processing applications such as image compressing[53], edge detection[54], Image Retrieval[12] and so on.

Feature selection is a method that finds and removes redundant and irrelevant feature components. As a result, low time complexity and high system accuracy could be achieved. Many feature selection approaches have been proposed for different applications such as face recognition [14], data mining and pattern recognition [16], automatic speaker verification system [50, 41-42] and image processing applications [39-40 and 46-47]. Recently ACO has been applied for feature selection task [3, 20]. Feature selection has also been applied to CBIR systems [11, 49].

In this paper, two feature categories including texture and color features are extracted in feature extraction phase. In texture features, the wavelet transform of image is computed and then the Frobenius norm of low frequency components are computed as texture features in decomposed images. For color features, dominant color descriptor is extracted from RGB and HSV image color domains for compact representative of image colors. Also, color statistic features and color histogram features are extracted as color features. The similarity measure must be adapted for feature types, so, we applied appropriate similarity measures for each feature type. Finally, the new feature selection schema based on ant colony optimization is proposed to select the most relevant features among texture and color features which lead the CBIR system with lower feature number and higher F-measure.

The rest of this paper is organized as follows: Section 2 describes the wavelet features. Color features are explained in more details in Section 3. The similarity measures for the proposed features are described in Section 4. Proposed feature selection algorithm based on ACO is presented in section 5. Also, Experimental setup and results are described in Sections 6 and 7 respectively. Finally, the conclusion and future works are discussed in Section 8.

## 2. Wavelet features

### 2. 1 Wavelet transformation

Wavelet transform could be applied to images as 2-dimensional signals. To refract an image into k level, first the transform is applied on all rows up to k level while columns of the image are kept unchanged. Then this task is applied on columns while keeping rows unchanged. In this manner frequency components of the image are obtained up to k level. These components are LL that is approximation of image and HL, LH and HH that are horizontal, vertical and diagonal frequency details [6].

These frequency components in various levels let us to better analyze original image or signal. Wavelet Transform in 2-D can be defined as following:

$$CWT(s,a,b) = \frac{1}{\sqrt{s}} \iint f(x,y)\, \Psi\left(\frac{x-a}{s}, \frac{y-b}{s}\right) dxdy \tag{1}$$

For a sample image, the wavelet decomposition for 2 levels is shown in Figure 1.

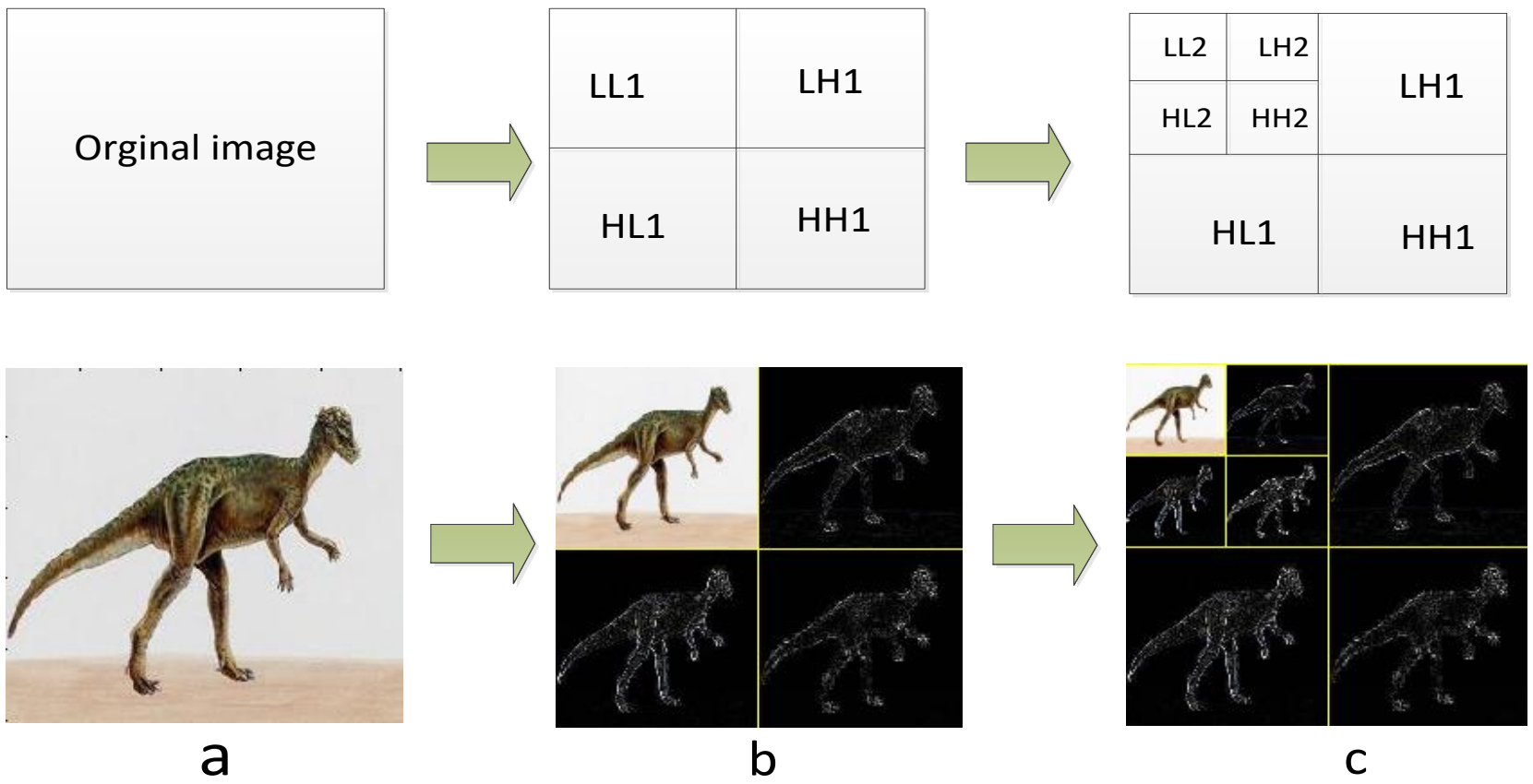

**Figure ١ . Wavelet decomposition of a sample image for 2 levels**

## 2.1. Wavelet Coefficient Features

Wavelet features are extracted based on the wavelet transformation which is described in the previous section. Here we extract features using approximation, vertical and horizontal frequency components of the image. The decomposed image consists of LL, LH, HL and HH components in each level. LL includes low frequency factors and HL, LH and HH include factors of high frequencies in horizontal, vertical and diagonal directions respectively. These frequency components are equal-sized matrices. The Frobenius norm of the rows of LL and LH matrices and also the Frobenius norm of the columns of HL matrix are proposed to be used in feature extraction phase as shown in Figure 2. The Frobenius norm of a vector is defined as following:

$$M \in R^n \implies \|M\|_F = (\sum_{i=1}^{n} (M_i)^2)^{\frac{1}{2}} \qquad (2)$$

In which $M = [M_1, M_2, \dots, M_n]$ is $n$ dimensional vector.

For images in RGB space, each color component (R,G and B) is decomposed to first level of Haar filter LL, LH and HL component. Then the vectors $a$, $h$ and $v$ are computed by the way that each element of $a$, $h$ and $v$ is the norm of each row in LL, and LH and each column of HL respectively. The standard deviation and mean of $a$, $h$ and $v$ create the features for images.

This method aims for reducing sensitivity toward local differences existing in each group of images in the data set such as changes in the state and position of objects in images. These differences can result in decreasing distances within a single group and increasing them between different groups.

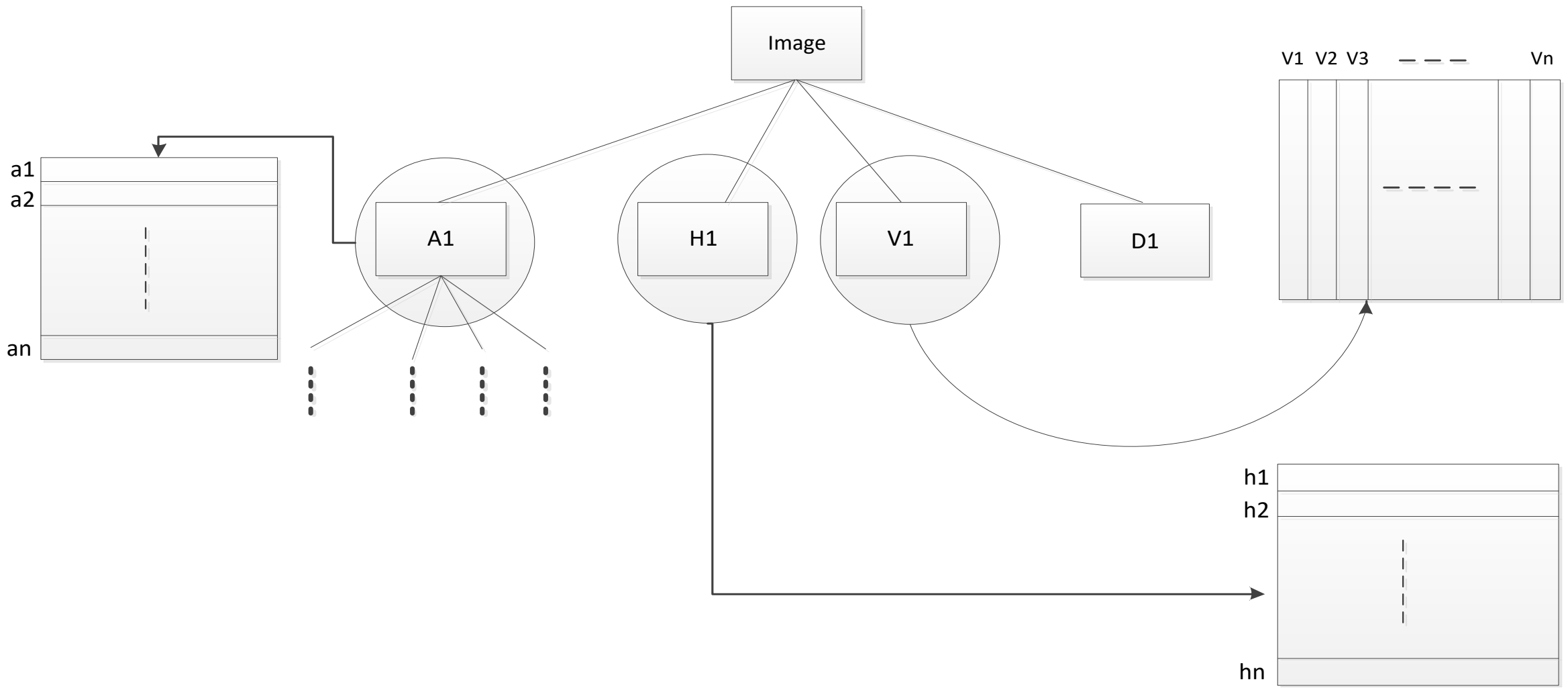

**Figure ٢ . Frequency components for feature extraction**

## 3. Color Features

Color features are the most intuitive and most dominant low-level image features which are very stable and robust in comparison with other image features such as texture and shape. Since, these features are not sensitive to rotation, translation and scale changes, they could be applicable for CBIR systems. What is more, the color feature calculation cost is relatively lower than other features. In this work we have used the following color-based features.

### 3.1. Dominant Color Descriptor (DCD)

DCD is one of the approved color descriptors in the MPEG-7 Final Committee Draft among several number of histogram descriptors. Both representative colors and the percentage of each color are included in DCD. Moreover, DCD provide an effective and compact color representation, could be applied for color distribution in an image or a region of interesting [15]. In DCD, first, each color is divided into the number of partitions named course partitions as depicted in Figure 5.

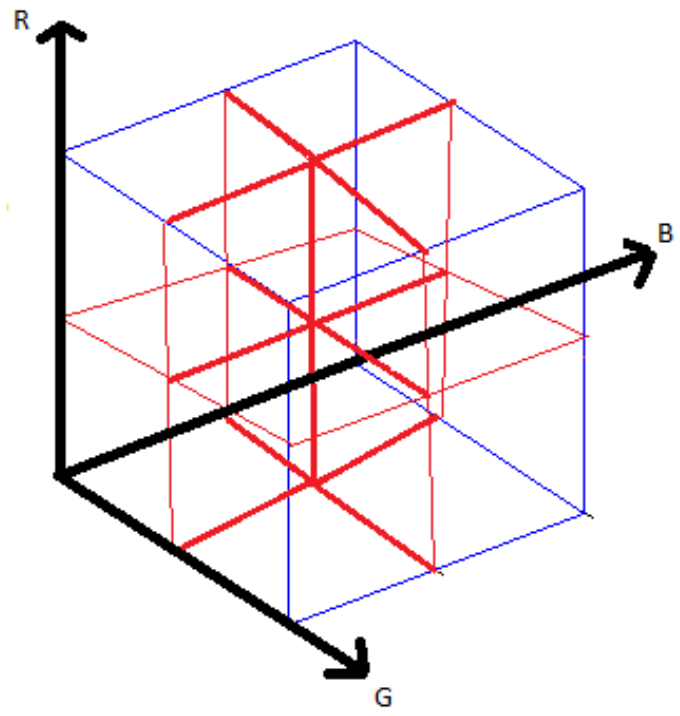

**Figure 3. Color space partitioning**

The center of each partition is calculated. Then, all points in a same partition are assumed to be similar and near to each other. Partition centers are the average value of all pixels in each partition and they are calculated as following:

$$C_j = \frac{\sum_{p \in P_i} p}{\sum_{p \in P_i} 1} \quad (3)$$

Which $P_i$ is a ith partition. In this research the DCD features are extracted in both RGB and HSV domains. Each pixel color value is replaced by the center value of its corresponding partition. As a result DCD quantizes the images which the possible colors for each pixel is equal with the partition numbers. Figures 6 and 7 show the result of applying DCD to image in RGB and HSV domains respectively.

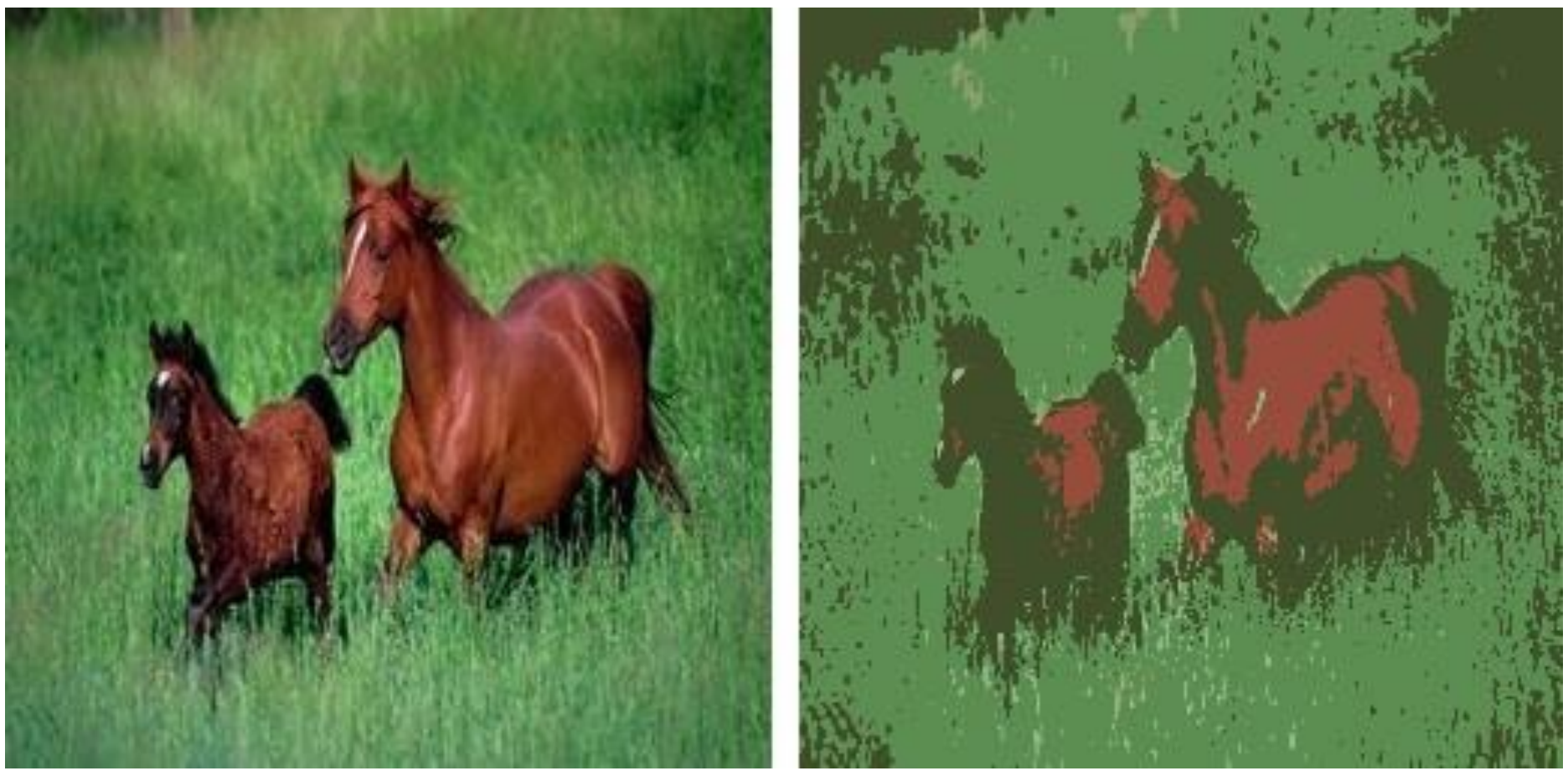

Figure. 4. DCD of the image in RGB color space

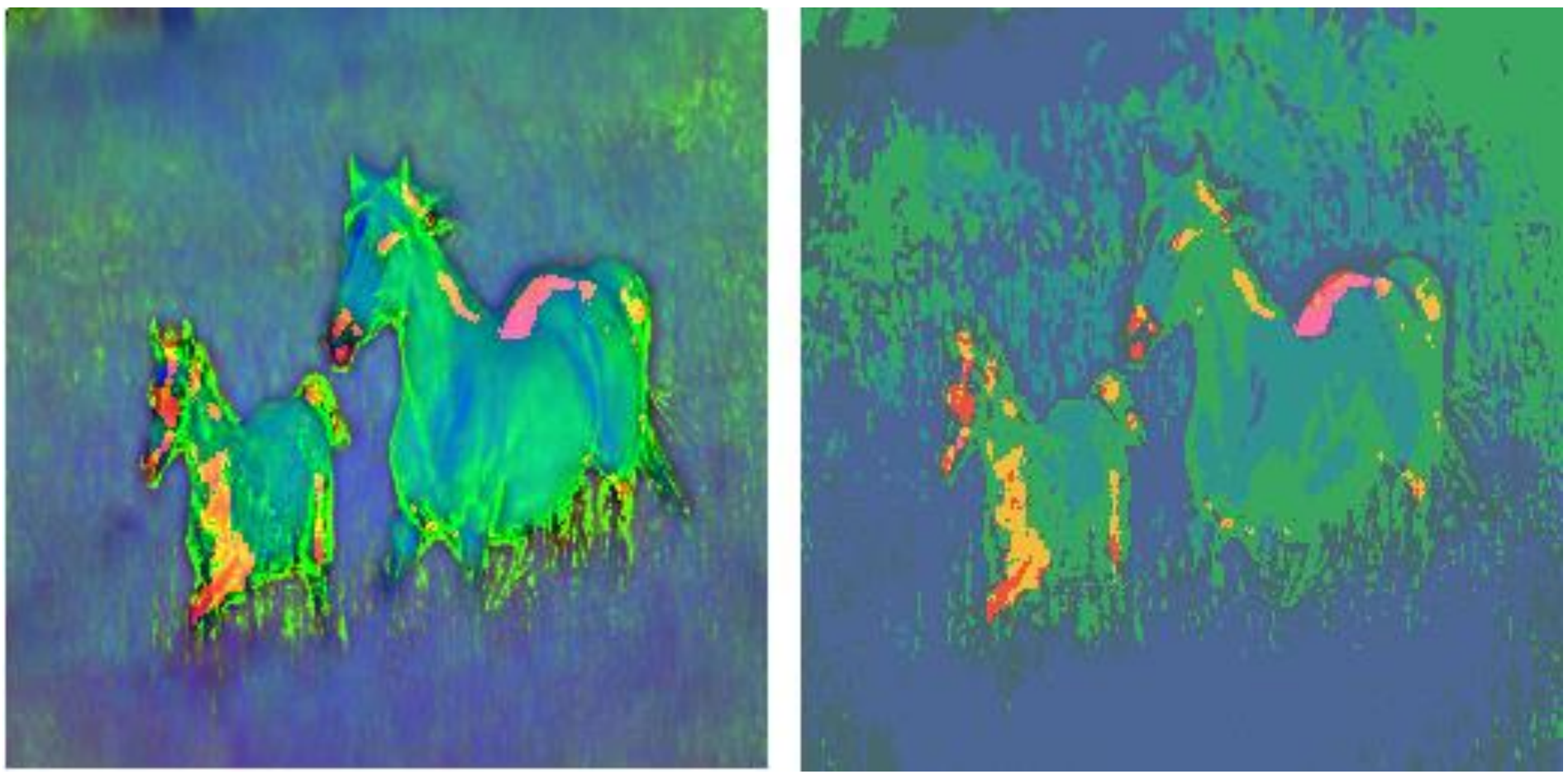

Figure 5. DCD of the image in HSV color space

### 3.2. Color statistic features

Usually images are used in the literature in both RGB and HSV space since these spaces are in bisection with one another. For each image, as the simplest statistic features, the first order (mean, denoted by M) and the second order (standard deviation, denoted by STD) are color statistics with respect to R, G, B, H, S and V channels and gray level image. As a result, fourteen features are generated per image with the following names:

$R_M, R_{STD}, G_M, G_{STD}, B_M, B_{STD}, H_M, H_{STD}, S_M, S_{STD}, V_M, V_{STD}, Gray_M \text{ and } Gray_{STD}.$

### 3.2 Color histogram features

The histogram is a graph which shows the number of color values falling in a number of resolution ranges or bins. From image histogram, a set of features could be extracted which are named as color histogram features [48]. For each color channel such as R, G, B, H, S and V a color histogram graph is drawn with respect to bin size B (the number of intervals or range resolutions). Each color channel has B histogram features corresponding to number of pixels which are dropped in each interval.

## 4. Similarity Measures

After feature extraction, images are retrieved based on their similarity with images in database. In this section the similarity measures for each type of features are proposed.

### 4.1. Texture and color statistic features similarity measure

As discussed in texture feature extraction, the standard deviation of special component of wavelet decomposition is computed as texture features. Also, the standard deviation and mean of color components are computed as color statistic features. Euclidean distance is used as similarity measure for both of these features.

$$D(Image1, Image2) = \sqrt{\sum_{i=1}^{n} (p_i - q_i)^2} \quad (4)$$

Where $p_i$ and $q_i$ are the ith correspondence features in *Image1* and *Image2* respectively.

### 4.2. Color histogram similarity measure

Color histograms, which record the frequencies of colors in the image is used as features for describing images in order to perform image retrieval. The following distance measure is applied for retrieving image among image databases [2].

$$D_{hist}(Image1, Image2) = 1 - \sum_{k=1}^{N} \min(H_1(k), H_2(k)) \quad (5)$$

Where $N$ is the number of bins in histogram and H(k) is a percentage of colors in bin number k.

### 4.3. DCD similarity measure

DCD features are centers (dominant color) of partitioned color space. As it is important that each color coarse partition contain how many pixels in it, the weighted Euclidean distance measure is presented for these features. It means that the distance of two correspondence centers is affected by percentage of pixels in that center.

$$D(Image1, Image2) = \sum_{i=1}^{n} w_i \sqrt{(C1_i - C2_i)^2} \quad (6)$$

Where $C1_i$ is the ith dominant color in Image 1 and $w_i = \frac{(p1_i + p2_i)}{2}$ . $p1_i$ is the percentage of pixels in ith center in *Image1* and sum of all $w_i$s is 1.

## 5. Proposed feature selection based on ACO

ACO is a system based on agents, which simulate the natural behavior of ants, consisting of mechanisms of adaptation and cooperation. The ability of ants to find shortest paths is achieved by their depositing of pheromone when they travel. Each ant probabilistically prefers to follow a direction that has more pheromone and the pheromone decays over time. Given that over time, the shortest paths will have more pheromone and higher chance for selection by ants. This path will be reinforced and the other paths' pheromone diminished until all ants follow the same path. In this way, the shortest path is achieved and system converges to a single solution [9].

ACO can be reformulated to solve feature selection problem. The main idea of ACO is to find a path with minimum cost in graph. ACO starts to search for the optimal feature subset with the ant's traverse through the graph until a minimum number of nodes are visited and traversal stop criterion is satisfied [28]. The graph is fully connected to allow any feature to be selected in the next stages. Based on this reformulation of the graph representation, the pheromone update rule and transition rule of standard ACO algorithm can be used. In this case, pheromone and heuristic value are not associated with edges. Instead, each feature has its own pheromone value and heuristic value because edges do not affect the optimum path but the features affect it. The probability that ant *k* selects feature *i* at time step *t* is [29]:

$$P_i^k(t) = \begin{cases} \dfrac{|\tau_i(t)|^{\gamma}|\eta_i|^{\delta}}{\sum_{u \in J^k}|\tau_u(t)|^{\gamma}|\eta_u|^{\delta}} & if\ i \in J^k \\ 0 & otherwise \end{cases} \tag{7}$$

where $J^k$ is the set of features that are allowed to be added to the partial solution if they are not visited so far. $\tau_i(t)$ and $\eta_i$ are the pheromone value and heuristic desirability associated with feature *i* respectively. $\gamma$ and $\delta$ are two parameters that determine the importance of the pheromone value and the heuristic information respectively [25]. After each ant has completed its tour, the solution is generated and then allowing each ant to deposit pheromone on the features that are part of its tour. The amount of pheromone deposited by ant $k$ on feature *i* in step *t* is proposed to be defined as following:

$$\Delta\tau(i,k) = \propto.(CBIR_F_measure(k)) + \beta.\left(\frac{FeaturesNumber - FeaturesNumber(k)}{FeaturesNumber}\right) \tag{8}$$

where $CBIR_F_measure(k)$ is a F-measure of CBIR system with the features which are found by ant $k$ at iteration *t*. $FeaturesNumber$ and $FeaturesNumber(k)$ are the number of all features and the number of features which are found by ant $k$ respectively, $\propto$ and $\beta$ are parameters that control importance of classifier performance and feature subset length respectively . After all ants have completed their solutions, the pheromone trails are updated by the following relation [28]:

$$\tau_i(t+1) = (1-\rho).\tau_i(t) + \sum_{k=1}^{m} \Delta\tau_i{}^k(t) + \Delta\tau_i{}^g(t \qquad (9)$$

where $\rho$ is an evaporation rate constant, $m$ is the number of ants and $g$ is the best ant in previous iteration. This relationship means that all the ants can update the pheromone and the one with the best solution deposits additional pheromone on nodes. This causes the search of ants to stay around the optimal solution in next iterations.

In this paper the stop criteria for ants is a threshold which is proposed to be defined as following:

$$\text{Ant_Threshold} = \varphi * \exp^{-\frac{\text{FN}}{\text{N}}} + \omega * \exp^{\text{F_Measure}} \qquad (10)$$

where $FN$ is the feature cardinality of the selected feature by the ant so far, $N$ is the number of all features, $\varphi$ and $\omega$ are the parameters that control the effect of feature size and F_measure respectively with restriction $\varphi + \omega = 1$.

We applied ACO-based feature selection in CBIR system. Ants select most relevant features based on their performance in CBIR system. Here, we use F-measure, a combination of precision and recall measures, for CBIR performance evaluation. So, the main advantage of presenting most relevant features for CBIR system is a system with the low complexity and high accuracy. The block diagram of proposed feature selection schema for CBIR systems is shown in Figure 6.

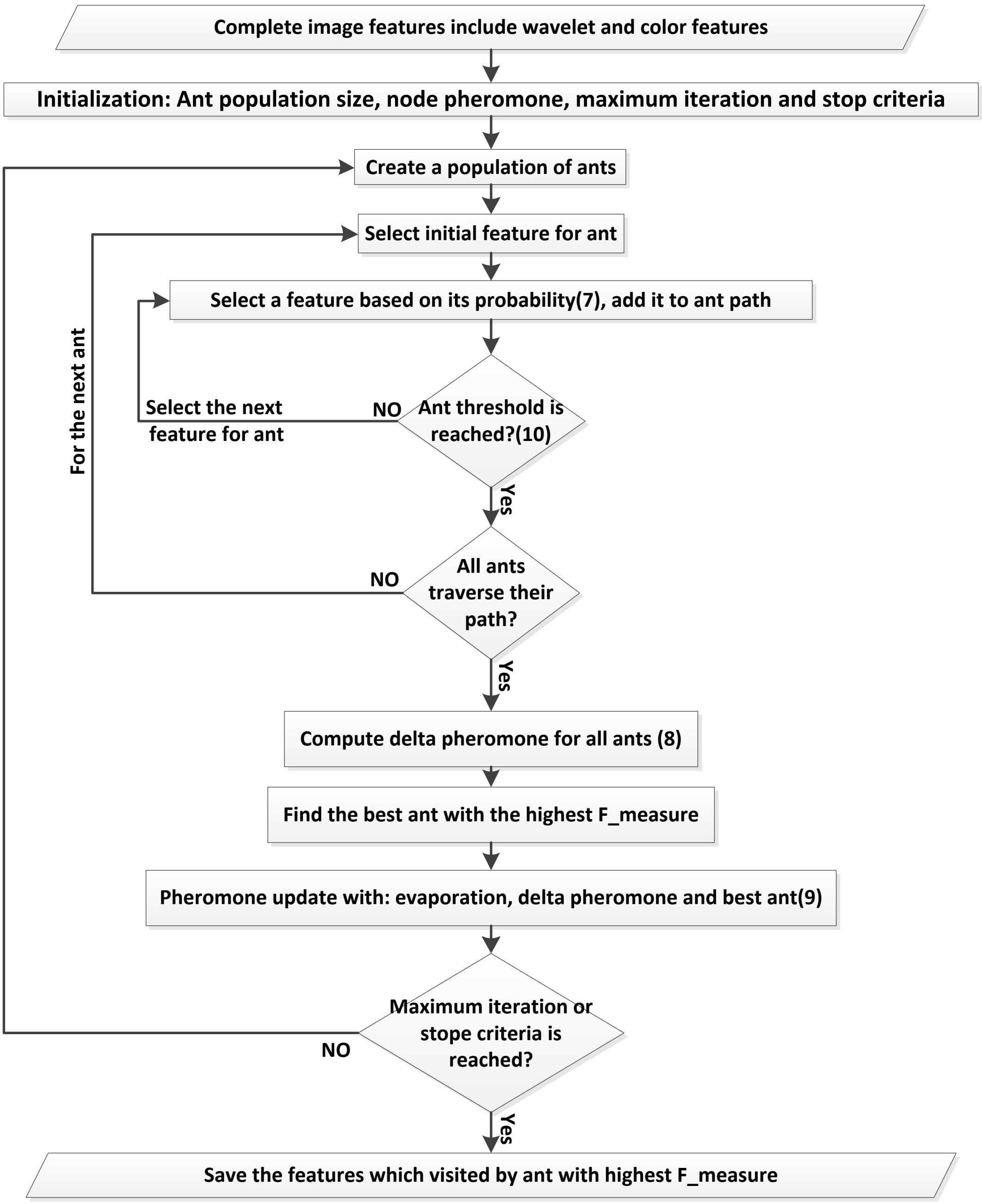

Figure 6. Proposed ACO feature selection schema in CBIR system

## 6. Experimental setup

To show the utility of the proposed CBIR schema, it is implemented on a machine with 3.4 GHz corei7 CPU, 6GB of RAM and windows 7. For simplicity, all images are resized to 256x256 pixels. For texture features, since the images are in the size of 256x256, the wavelet decompositions LL, LH and HL are all 128x128. The norm of each row in LL and LH and the norm of each column in HL are all the vector of size 1x128. So, for each color component we have 3 features as standard deviation and 3 features as mean of these three vectors. Since just the RGB color space is used for wavelet features, the feature vector of size 18 is computed per image. The number of dominant colors in DCD feature extraction is 8. So 48 features are extracted as DCD features in R, G, B, H, S and V color components. Furthermore, 12 features are extracted as standard deviation and mean of 6 color space in RGB and HSV. Finally, in histogram features we have set the bin number to 8. Therefore, for RGB and HSV 48 features could be extracted as histogram features. As a result, after extraction all of mentioned features, for each image 126 features are extracted. Table 1 shows the extracted features from image.

Table 1. Abstract name of all extracted features

| *Feature Sample* | *Feature Meaning* | *Feature Sample* | *Feature Meaning* | *Feature Sample* | *Feature Meaning* |
|---|---|---|---|---|---|
| LL_meanL | Mean of norm of rows in LL | $\mathrm{DCDS}_i$ | ith color dominant of S | $\boldsymbol{V_m}$ | Mean of V component |
| LH_meanL | Mean of norm of L Components of LH | $\mathrm{DCDV}_i$ | ith color dominant of V | $\boldsymbol{H_s}$ | Std of H component |
| HL_meanL | Mean of norm of L Components of HL | $\boldsymbol{R_m}$ | Mean of R component | $\boldsymbol{S_s}$ | Std of S component |
| LL_stdL | Std of norm of rows in LL | $\boldsymbol{G_m}$ | Mean of G component | $\boldsymbol{V_s}$ | Std of V component |
| LH_stdL | Std of norm of L Components of LH | $\boldsymbol{B_m}$ | Mean of B component | $\mathrm{HistR}_i$ | ith color bin of R |
| HL_stdL | Std of norm of L Components of HL | $\boldsymbol{R_s}$ | Std of R component | $\mathrm{HistG}_i$ | ith color bin of G |
| $\mathrm{DCDR}_i$ | ith color dominant of R | $\boldsymbol{G_s}$ | Std of G component | $\mathrm{HistB}_i$ | ith color bin of B |
| $\mathrm{DCDG}_i$ | ith color dominant of G | $\boldsymbol{B_s}$ | Std of B component | $\mathrm{HistH}_i$ | ith color bin of H |
| $\mathrm{DCDB}_i$ | ith color dominant of B | $\boldsymbol{H_m}$ | Mean of H component | $\mathrm{HistS}_i$ | ith color bin of S |
| $\mathrm{DCDH}_i$ | ith color dominant of H | $\boldsymbol{S_m}$ | Mean of S component | $\mathrm{HistV}_i$ | ith color bin of V |

Each wavelet feature in Table 1 is extracted for all 6 color components R, G, B, H, S and V. Finally, the parameters for ACO-based feature selection are set as:

$\gamma = 1, \delta = 2, \propto = 0.7, \beta = 0.3, \rho = 0.2, \varphi = 0.2$ and $\omega = 0.8$.

## 7. Experimental results

Our proposed system is evaluated with the Corel image database which have been widely used by the image processing and CBIR research communities that consists of 11,000 images of 110 categories which the test classes in our evaluation are including bus, horse, flower, dinosaur, building, elephant, people, beach, scenery and dish. In this paper, first, the ACO feature selection is adapted for CBIR system which gives the most relevant features among all features extracted from images. In this experiment ACO feature selection selects 42 top features among all 126 features which lead to 66% of feature reduction. Although ACO feature selection is a time-consuming task, it is an offline task and performed once. Selected features by feature selection algorithm are LL_meanL_G, LL_meanL_B, LH_meanL_B, HL_meanL_R, LL_stdL_G, LH_stdL_R, HL_stdL_R, HL_stdL_B, DCDR4, DCDR7, DCDG2, DCDB4, DCDB6, DCDB7, DCDH3, DCDH8, DCDS5, DCDV5, DCDV6, $\mathrm{R_m}$, $\mathrm{R_s}$, $\mathrm{B_s}$, $\mathrm{H_m}$, $\mathrm{S_m}$, $\mathrm{H_s}$, $\mathrm{V_s}$, HistR3, HistR8, HistG1, HistG5, HistG8, HistB1, HistB2, HistH6, HistH7, HistH8, HistS2, HistS7, HistV4, HistV5, HistV7 and HistV8.

After feature selection, the indexes of selected features are presented for proposed CBIR system which is shown in Figure 7.

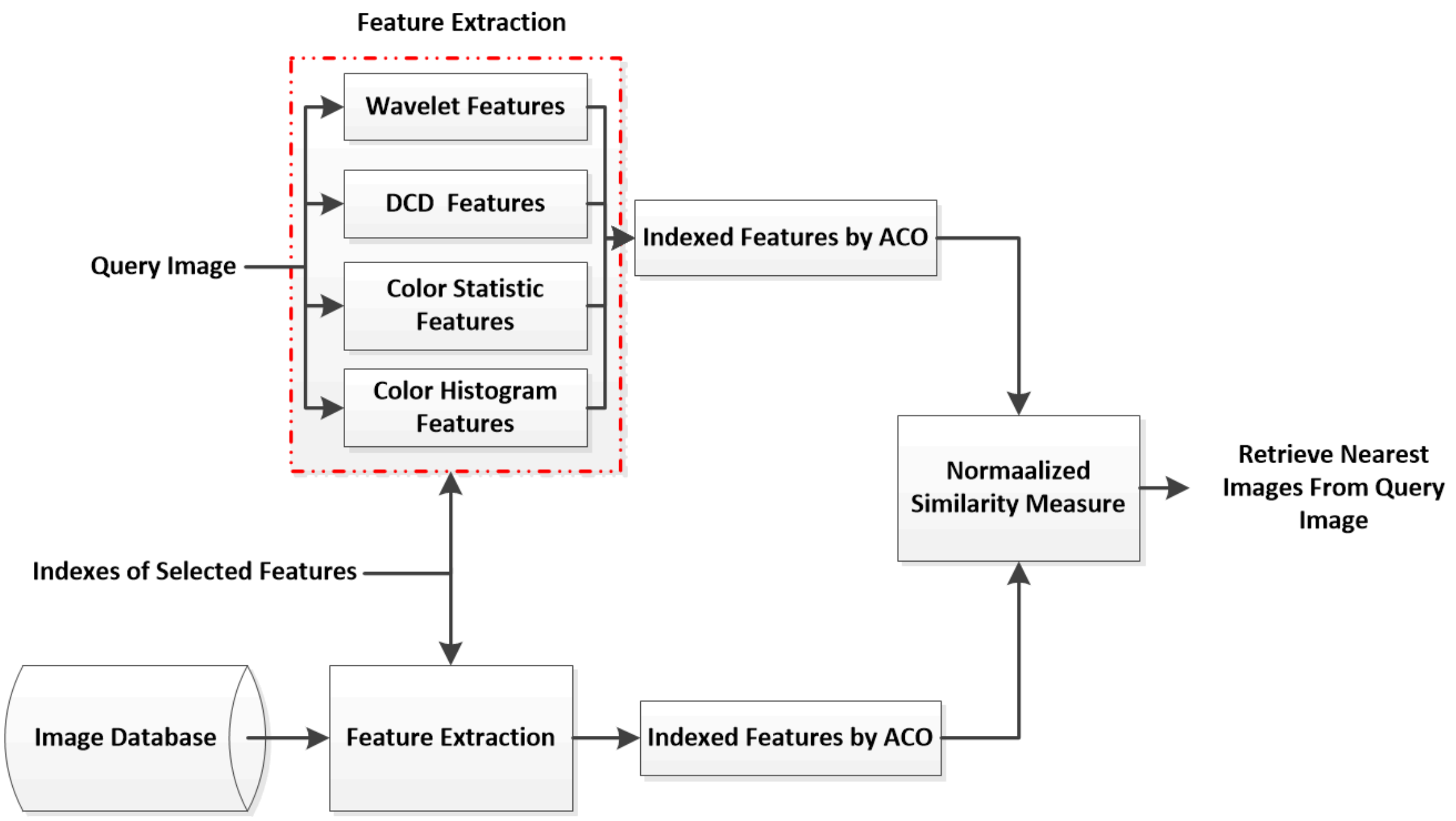

**Figure 7. Proposed CBIR system**

To evaluate the performance of the proposed system, it has been compared with CBIR system using color, texture and shape features [38], co-occurrence matrix-based CBIR[24] and Image retrieval by texture similarity[30]. From Corel database, all images have been used as query images (the tested 10 semantic class includes bus, horse, flower, dinosaur, building, elephant, people, beach, scenery, and dish) and then the first 20 most similar images are retrieved. For each class of image, the both average normal precision and recall are computed for all 100 query images in each class. Results for all image categories are show in Table 2.

Table 2. Average precision and recall for proposed CBIR system and other systems

| Image Category | Proposed System | | CBIR with color, texture and shape features[38] | | co-occurrence matrix-based CBIR[24] | | Image retrieval by texture similarity[30] | |
|---|---|---|---|---|---|---|---|---|
| | Precision | Recall | Precision | Recall | Precision | Recall | Precision | Recall |
| African people | 0.583 | 0.116 | 0.522 | 0.104 | 0.453 | 0.115 | 0.424 | 0.126 |
| Beach | 0.489 | 0.097 | 0.462 | 0.092 | 0.398 | 0.121 | 0.446 | 0.113 |
| Building | 0.455 | 0.091 | 0.444 | 0.088 | 0.374 | 0.127 | 0.411 | 0.132 |
| Buses | 0.546 | 0.109 | 0.558 | 0.111 | 0.741 | 0.092 | 0.852 | 0.099 |
| Dinosaurs | 0.998 | 0.199 | 0.892 | 0.178 | 0.915 | 0.072 | 0.587 | 0.104 |
| Elephants | 0.469 | 0.093 | 0.442 | 0.088 | 0.304 | 0.132 | 0.426 | 0.119 |
| Flowers | 0.896 | 0.179 | 0.863 | 0.172 | 0.852 | 0.087 | 0.898 | 0.093 |
| Horses | 0.612 | 0.122 | 0.601 | 0.120 | 0.568 | 0.102 | 0.589 | 0.103 |
| Mountains | 0.398 | 0.079 | 0.446 | 0.089 | 0.293 | 0.135 | 0.268 | 0.152 |
| Food | 0.633 | 0.126 | 0.619 | 0.123 | 0.369 | 0.129 | 0.427 | 0.122 |
| **Average** | **0.6079** | **0.1211** | **0.5849** | **0.1165** | **0.527** | **0.111** | **0.533** | **0.116** |

It is clear that the proposed CBIR system has achieved better average precision and recall of various image categories than the other three models. The sample result for Image retrieval by texture similarity, co-occurrence matrix-based CBIR, CBIR system using color, texture and shape features and our proposed system are shown in Figures 8, 9, 10 and 11 respectively. The image at the top of left corner is the query image and other 20 images are the retrieval results.

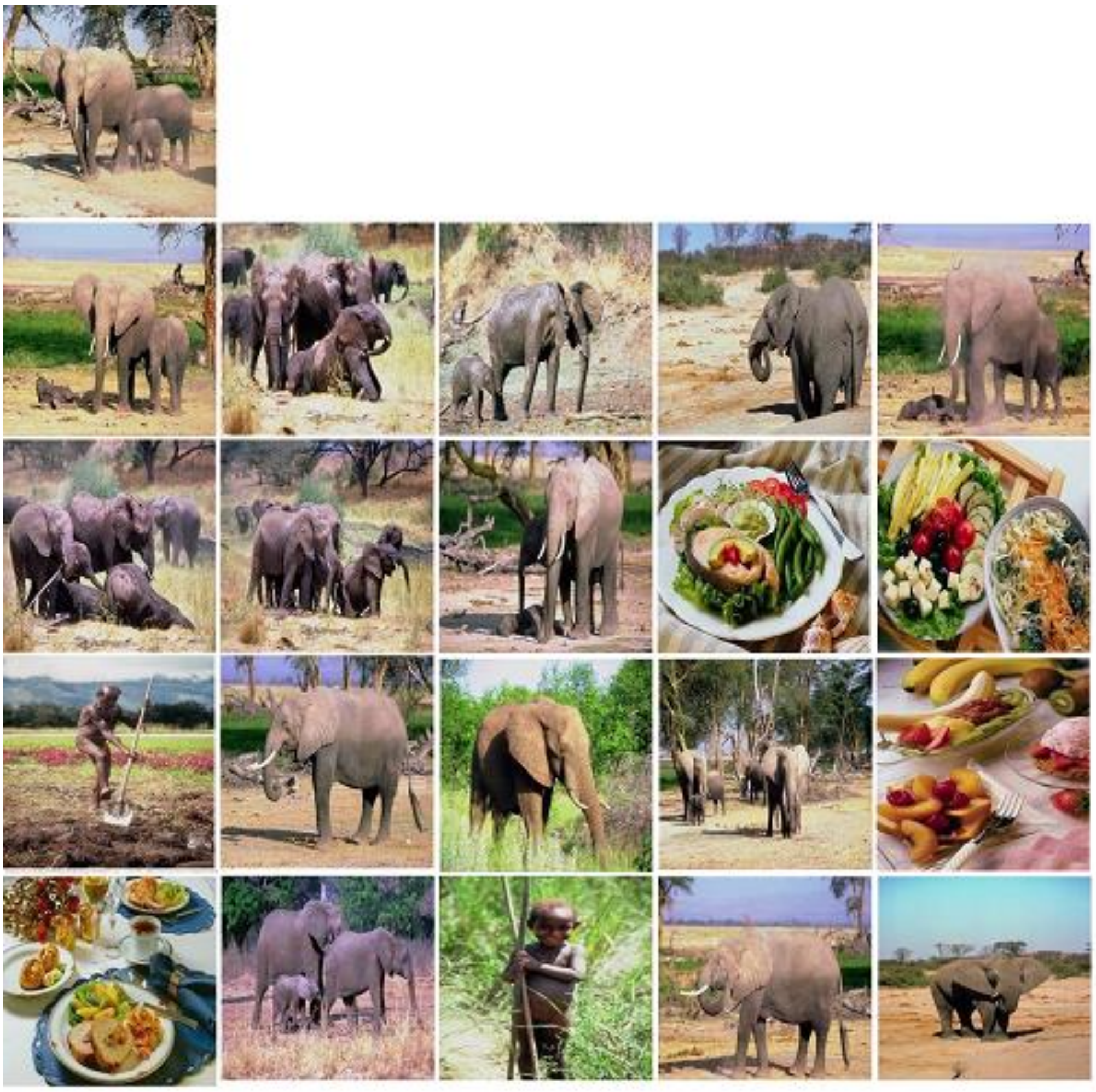

Figure 8. Retrived images for Image retrieval by texture similarity

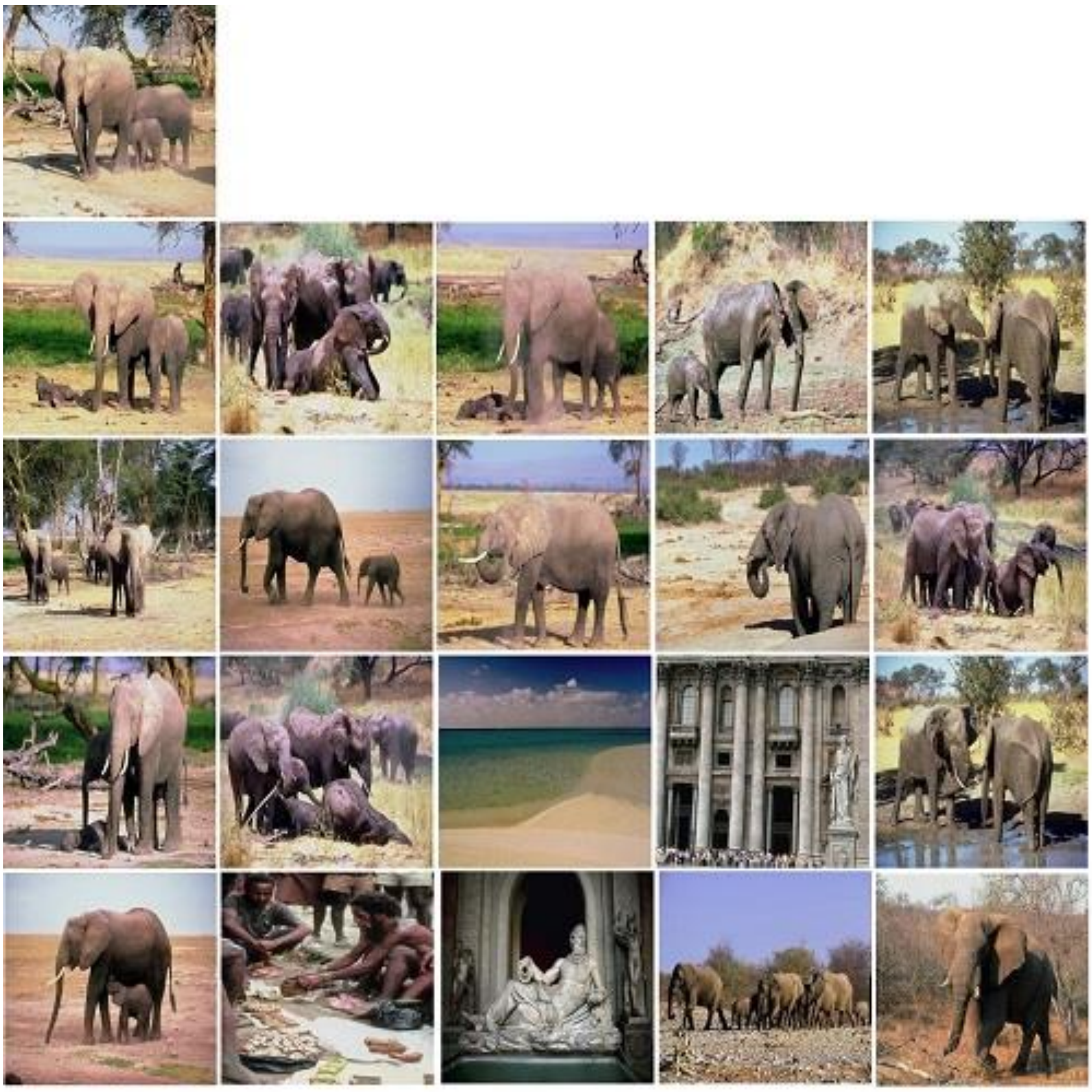

Figure 9. Retrived images for co-occurrence matrix-based CBIR

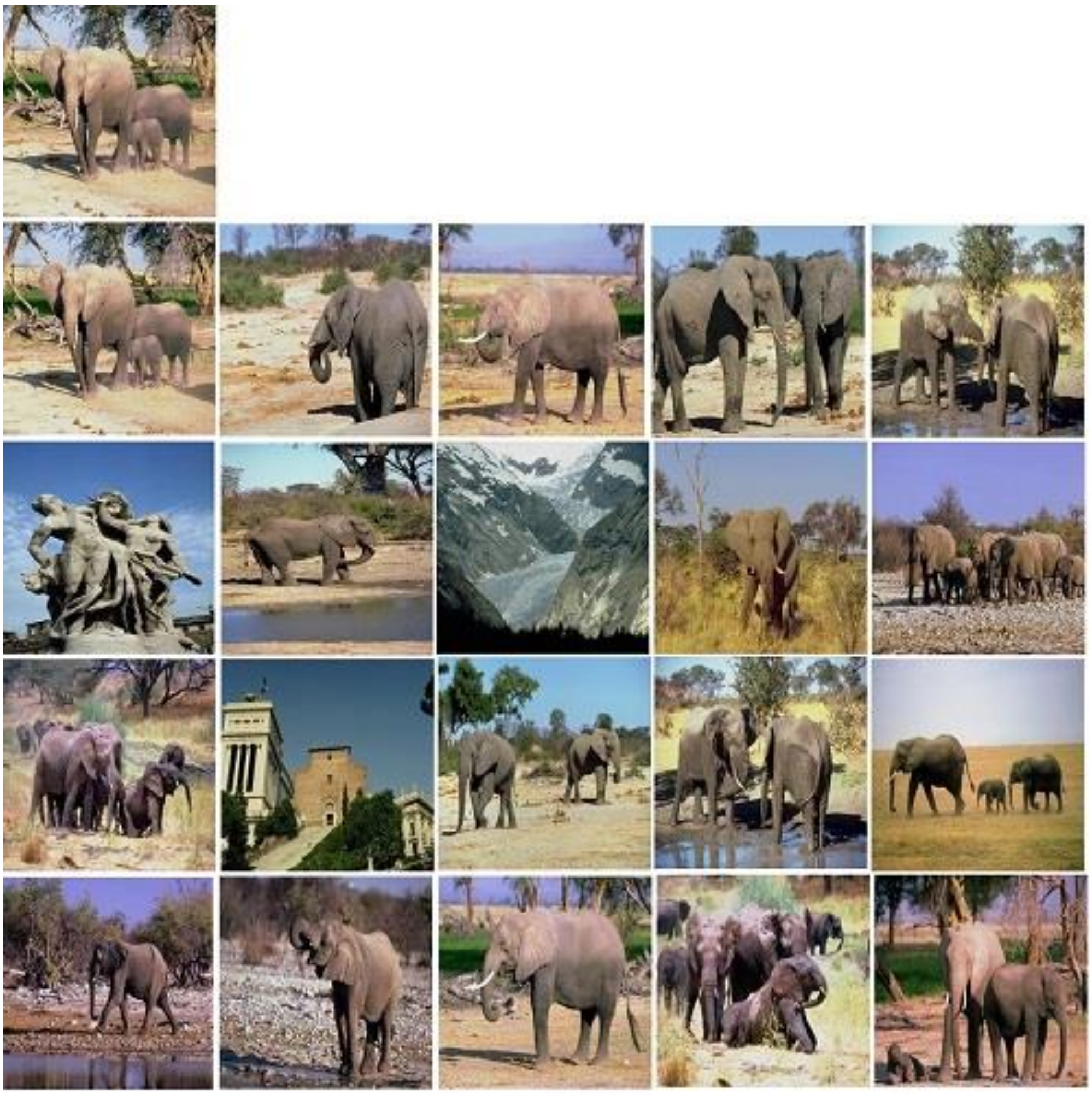

Figure 10. Retrived images CBIR system using color, texture and shape features

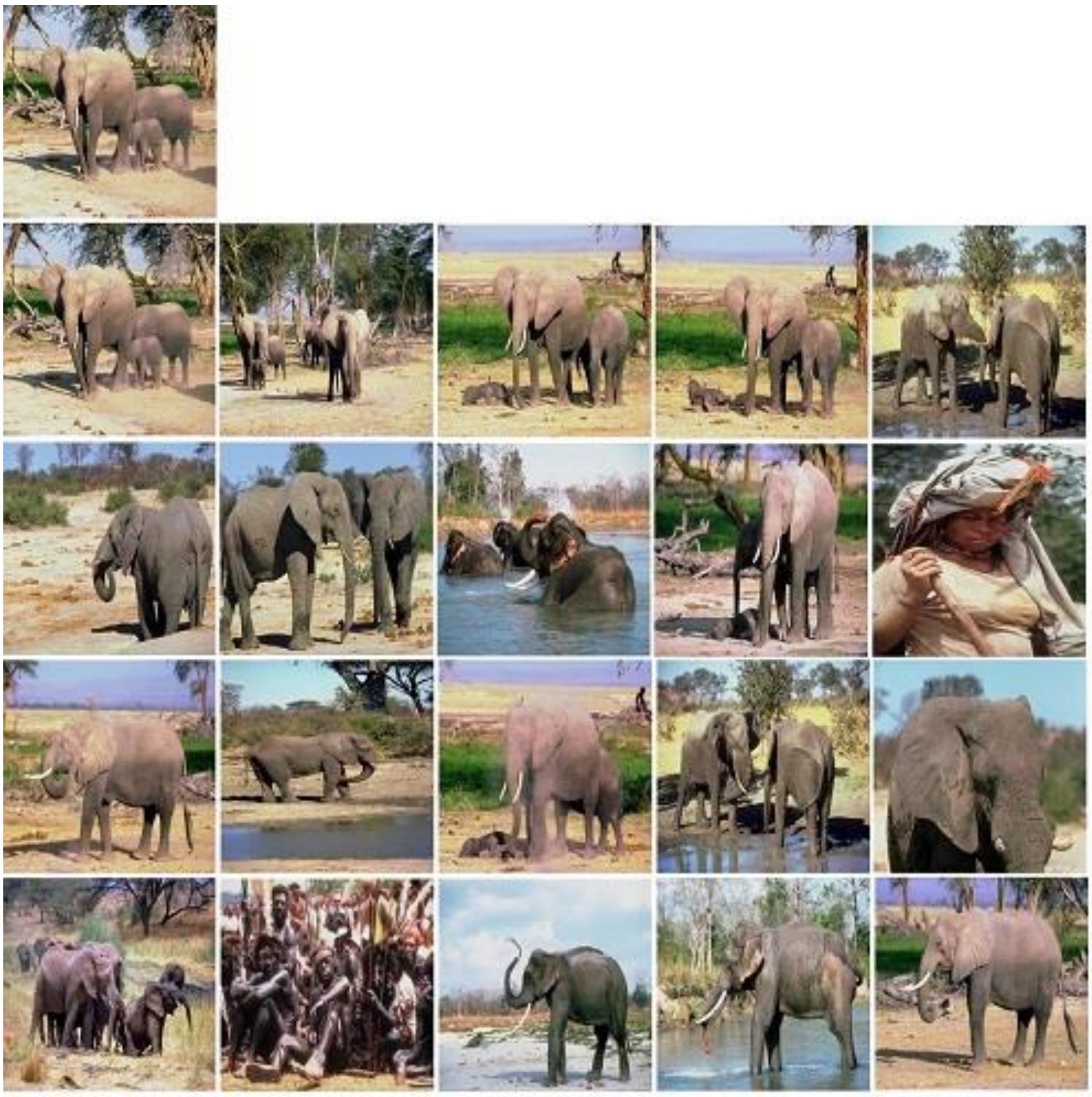

Figure 11. Retrieved images for proposed CBIR system

## 8. Conclusion

In this paper, the Frobenius norm of low frequency components of decomposed images from wavelet transform are proposed as wavelet features as well as proposed color features include DCD, color statistic and color histogram features. Feature extraction phase is developed by the way that distance between features of same image class is minimum while it is maximum for images of different classes which it leads to a CBIR system with higher performance. Also, all irrelevant and redundant features are dropped by ant colony optimization which selects most relevant features among complete feature set. Results showed that our proposed CBIR system has higher precision and recall in comparison with three older CBIR systems. For future works, curve let transform could be used for extracting features in higher frequency resolution rather than just low and high frequency components in wavelet transform. In addition, other feature selection methods such as particle swarm optimization and genetic algorithm can be applied and intrinsic property of data such as Relieff weights can be used in these feature selection methods to increase CBIR system performance.